Title: Reaction rates in squeezed polaron bands controlled by quantum statistics

Authors: Mladen Georgiev (1) and Alexander Gochev (2) (Institute of Solid State Physics, Bulgarian Academy of Sciences, 1784 Sofia, Bulgaria (1), PGA Solutions, Galloway, OH 43119, U.S.A. (2))

Reaction rates are usually defined using classical statistics for introducing the thermal occupation probabilities. Its predictions for the temperature dependence of a rate are found in reasonable agreement with experiments. In view of the applications to polaronic systems at lower temperatures under strongly quantized conditions, we now extend the definition so as to incorporate quantum statistics, as well, Fermi-Dirac for polarons and Bose-Einstein for bipolarons. We find both extensions feasible.

## 1. Introduction

Among the theories dealing with the phonon-coupled transition rates, whether electronic or ionic, the central position has been preserved for the reaction rate theories [1,2]. They are transparent physically and easily tractable [3,4] which makes them readily applicable to interpreting specific experimental data [5,6]. In Christov's definition, the medium is assumed to be a manifold of harmonic oscillators vibrating at the same bare Einstein frequency $\nu$ [7]. This approach has been found useful for dealing with the electronic relaxation at local (color) centers [8]. The definition has later been extended so as to cover the local rotation of off-centered ions [9]. The rotating entity proven to behaving as a nonlinear oscillator, its exact quantum-mechanical eigenstates are given by Mathieu's periodic functions, while the eigenvalues formed rotational bands [10].

Whether in linear- or in nonlinear-oscillator terms, we follow Christov's line to define the reaction rate by way of

$$k_{12}(T) = (k_B T / h) \sum_\alpha W(E_\alpha) P(E_\alpha) (\Delta E_\alpha / k_B T) \quad (1)$$

accounting for the elastic (energy-conserving) transitions at the total energy levels $E_n$ with transition probabilities $W(E_\alpha)$ for transfer across the energy barrier, $h$ and $k_B$ are Planck's and Boltzmann's constants, respectively. We introduce N for the number of particles and $E_\alpha$ for the total energies of the N- particle system and define them by

$$N = \sum_k n_k \quad (2)$$

$$E_\alpha = \sum_k n_k \varepsilon_k , \quad (3)$$

respectively, where $n_k$ are the occupation numbers, $\varepsilon_k$ are the single-particle quantized energy levels, $\alpha$ is the particle configuration $\alpha = \{n_k\}$. The quantity

$$P(E_\alpha) = (1 / Z) \exp(-E_\alpha / k_B T) \qquad (4)$$

is the probability that at temperature T the system have an energy $E_\alpha$ within the interval $\Delta E_\alpha$, Z is the partition function. From the normalization requirement $\sum_n P(E_n) = 1$,

$$Z = \sum_n \exp(-E_n / k_B T) \qquad (5)$$

The definition (1) presumes that the configurational relaxation through coupling to the phonon bath is sufficiently fast; if so (1) is the rate of the bottle-neck process.

For example, a harmonic-oscillator system is specified by $\varepsilon_k = (n_k + ½)\, h\nu$ so that summing up over the vibrational modes gives

$$Z = \exp(-½\, h\nu / k_B T)\, [1 - \exp(-h\nu / k_B T)]^{-1} = 1 / 2\sinh(h\nu / 2k_B T) \qquad (6)$$

For the rotational mode of nonlinear oscillators, $\varepsilon_k$ is the spectrum in a rotational band.

Equations (1) through (3) apply to any quantum system. The definition (1) of a rate is quite general and it does not specify the statistics. For most systems studied so far the statistics has been irrelevant, though there have been some instances where incorporating either Fermi-Dirac (F-D) or Bose-Einstein (B-E) statistics would have been desirable. Such is the theory of bond polarons occuring in the axial transport in high-$T_c$ superconductors in which pairing has been suppressed by pulsed magnetic fields or impurities to extend the "normal state" to much lower temperatures [11]. In addition to temperatures largely inferior to $T_c$, quantum statistics should have been preferred since these bond polarons appeared to be strongly quantized systems as well. On the other hand, pairing in squeezed bands whether as Cooper pairs or as real space bipolarons would invoke B-E statistics. All this necessitated considering a redefinition of the reaction rate (1) to describe F-D or B-E particles in polaron or polaron-like bands.

## 2. Quantum statistics

From the definition of a free energy we have $Z = \exp(-F / k_B T)$ so that

$$P(E_\alpha) = \exp[(F - E_\alpha) / k_B T] \qquad (7)$$

This suggests defining the probability for finding an oscillator in state α by means of the thermodynamic potentials. Actually, equation (7) does so for a canonical ensemble. In order to adapt the theory to polarons in squeezed bands, we will have to extend it so as to cover the grand canonical ensemble. For the latter ensemble, following Blatt [12],

$$P(E_\alpha)_{N\alpha} = \exp[(\Omega + \mu N - E_\alpha) / k_B T] \qquad (8)$$

Here Ω is the thermodynamic potential, μ is the chemical potential of the polaron gas. The suffix $_{N\alpha}$ implies that equation (8) relates to a given number of particles N within an ensemble $\alpha$. Summing up we get

$$\sum_{N\alpha} \exp[(\Omega + \mu N - E_\alpha) / k_B T] = 1 \qquad (9)$$

wherefrom we obtain the thermodynamic potential in the form

$$\exp(-\Omega / k_B T) = \sum_{N\alpha} \exp[(\mu N - E_\alpha) / k_B T] \qquad (10)$$

Inserting the number of particles (2) and total energy (3) following certain manipulations, we get eq. (8) factorized into equivalent terms each one for a given quantum state k:

$$\exp(-\Omega / k_B T) = \prod_k \sum_{nk} (u_k)^{nk} \qquad (11)$$

$$u_k = \exp[(\mu - \varepsilon_k)/k_B T] \qquad (12)$$

## 2.1. Fermi-Dirac statistics

For F-D statistics it is essential that the occupation numbers are either 0 or 1. For that reason, $\sum_{nk} (u_k)^{nk} = 1 + u_k$ and

$$\exp(-\Omega / k_B T) = \prod_k (1 + u_k) \qquad (13)$$

wherefrom we obtain for the F-D distribution

$$P(E_\alpha)_{N\alpha\ F\text{-}D} = \exp(\Omega / k_B T) \prod_k (u_k)^{nk} = \prod_k [(u_k)^{nk} / (1 + u_k)] \qquad (14)$$

in terms of the occupation numbers $n_k$. The probability function (14) is normalized to unity $\sum P(E_\alpha)_{N\alpha} = 1$ as verified easily. Substituting for $u_k$ from eq. (12) we also get

$$P(E_\alpha)_{N\alpha\ F\text{-}D} = \prod_k \{(\exp[(\mu - \varepsilon_k)/k_B T])^{nk} / (1 + \exp[(\mu - \varepsilon_k)/k_B T])\} \qquad (15)$$

The familiar textbook result obtains as an average occupation number:

$$\langle n_k \rangle = \sum_{nk} n_k P(E_\alpha)_{N\alpha} = u_k / (1 + u_k) = 1 / ( \exp[(\varepsilon_k - \mu) / k_B T] + 1 ) \qquad (16)$$

### 2.1.1. Derivative classical statistics

At $\varepsilon_k > \mu$ and higher temperatures $k_B T \gg \varepsilon_k - \mu$ the quantity $u_k = \exp[(\mu - \varepsilon_k) / k_B T] \ll 1$ so that

$$P(E_\alpha)_{N\alpha\ Classic} \sim \prod_k(\exp[(\mu - \varepsilon_k)/k_B T])^{nk} \qquad (17)$$

$$\langle n_k \rangle \sim u_k \equiv \exp[- (\varepsilon_k - \mu) / k_B T], \qquad (18)$$

The resulting statistics is called ‘classical’; actually this is a Boltzmann tail statistics appearing to control the average particle distribution at higher temperatures. Using (15) and (16) we define a classical thermal occupation probability by means of

$$P(\varepsilon_n)_{Classic} = \exp(-\varepsilon_n / k_BT) / \sum_n \exp(-\varepsilon_n / k_BT) \qquad (19)$$

Note the difference between eqs. (3) and (19) in the definitions of the energies $E_\alpha$ and $\varepsilon_n$. Boltzmann Tail (BT) classical statistics has been applied to the polaron gas at not too low temperatures. It has been argued that the chemical potential $\mu$ (Fermi's energy $\varepsilon_F$) does not enter explicitly for BT polarons, as it does not in eq. (19) [13]. Certainly, a BT polaron gas is an approximation in which the Fermi energy cancels out in the numerator and the denominator of the thermal occupation probability.

## 2.2. Bose-Einstein statistics

For B-E statistics occupation numbers are all nonnegative integers. Now from eqs. (11) and (12), we have

$$\exp(-\Omega / k_BT) = \prod_k [1 / (1- u_k)] \qquad (20)$$

and, consequently,

$$P(E_\alpha)_{N\alpha\ B\text{-}E} = \exp(\Omega / k_BT) \prod_k (u_k)^{nk} = \prod_k (u_k)^{nk} (1 - u_k) \qquad (21)$$

The probability function (21) is also normalized to unity $\sum P(E_\alpha)_{N\alpha} = 1$. We substitute for $u_k$ from eq. (12) to get, alternatively,

$$P(E_\alpha)_{N\alpha\ B\text{-}E} = \prod_k \{\exp[(\mu - \varepsilon_k)/k_BT]\}^{nk} \{1 - \exp[(\mu - \varepsilon_k)/k_BT]\} \qquad (22)$$

Again, the result familiar from textbooks obtains as an average occupation number

$$<n_k> = \sum_{nk} n_k P(E_n)_{N\alpha} = u_k / (1 - u_k) = 1 / ( \exp[(\varepsilon_k - \mu) / k_BT] - 1 ) \qquad (23)$$

## 3. Transition probabilities

Following Christov, the nuclear- (configurational-) tunneling probability $W_{conf}(E_n,E_n)$ for a horizontal isoenergetic transition conserving the phonon number can be calculated using the extension of a formula originating from Bardeen [2,3]:

$$W_{conf}(E_n)= 4\pi^2 | U_{if}(E_n) |^2 \sigma_i(E_n)\sigma_f(E_n) \qquad (24)$$

where

$$U_{if}(E_n) = - (\eta^2/2m)\{\chi_i(E_n) [d\chi_f^*(E_n)/dq] - \chi_f(E_n) [d\chi_i^*(E_n)/dq]\} |_{q=qC} \qquad (25)$$

$U_{if}(E_n) = -i\eta j_{fi}$ is the potential induced by the transition current $j_{fi}$ at crossover. Here $\sigma_i(E_n)$ and $\sigma_f(E_n)$ are the DOS, $\chi_i(E_n)$, $\chi_f(E_n)$ are the nuclear-oscillator wave functions in the initial and final electronic states, respectively, $E_n$ is the energy of the vibronic transition. Using harmonic-oscillator wave functions normalized in Q-space:

$$\chi_n(q)_{\pm} = [\sqrt{(\alpha/\pi)}/2^n n!]^{½} H_n(q)\exp(-(q\pm q_0)^2/2) \qquad (26)$$

where $\alpha = M\omega^2/\eta\omega$, $q = \sqrt{\alpha}\, Q$ is the scaled and Q the actual configurational coordinate, $q_0$ is the absolute position along q of the well bottom, $q_C$ is the crossover coordinate. $H_n(q)$ are Hermite polynomials, $E_n = (n+½)\eta\omega$, $\sigma_i(E_n) = \sigma_f(E_n) = 1/\eta\omega$.

The elastic nuclear-tunneling probability at any finite $\Theta_p = p\hbar\omega$, the reaction heat at 0 K, nonpositive or positive, reads

$$W_{conf}(E_n,E_m) = \pi\{[F_{nm}(\xi_0,\xi_C)]^2/2^{n+m}\, n!m!\}\ \exp(-\varepsilon_R/\eta\omega)\ \exp(-\Theta_p^2/\eta\omega\varepsilon_R), \qquad (27)$$

where the vibronic level number in final electronic state is $m = n + p$ (at $\Theta_p < 0$) and $m = n - p$ (at $\Theta_p > 0$), or equivalently,

$$W_{conf}(E_n,\Theta_p) = \pi\{[F_{n,n\pm p}(\xi_0,\xi_C)]^2/2^{2n\pm p}\, n!(n\pm p)!\}\ \exp(-\varepsilon_R/\eta\omega)\ \exp(-\Theta_p^2/\eta\omega\varepsilon_R)$$

Using harmonic-oscillator wave functions for $\xi = q - 2q_0$:

$$F_{nm}(\xi_0,\xi_C) = \xi_0 H_n(\xi_C)H_m(\xi_C-\xi_0) - 2nH_{n-1}(\xi_C)H_m(\xi_C-\xi_0) +$$

$$2mH_n(\xi_C)H_{m-1}(\xi_C-\xi_0), \qquad (28)$$

or equivalently,

$$F_{n,n\pm p}(\xi_0,\xi_C) = \xi_0 H_n(\xi_C)H_{n\pm p}(\xi_C-\xi_0) - 2nH_{n-1}(\xi_C)H_{n\pm p}(\xi_C-\xi_0) +$$

$$2(n\pm p)\, H_n(\xi_C)H_{n\pm p-1}(\xi_C-\xi_0)$$

Here the parameters are

$$\varepsilon_C \equiv ½\, K\, Q_C^2 = ½\, \eta\omega\, q_C^2 = (\varepsilon_R + \Theta_p)^2 / 4\varepsilon_R \qquad (29)$$

is the crossover energy,

$$\varepsilon_R \equiv\ 2 \times ½\, KQ_0^2 = KQ_0^2 = \eta\omega\, q_0^2 \qquad (30)$$

is the lattice reorganization energy, K is the stiffness. In so far as $q_0$ is p-independent, so is $\varepsilon_R$. $V_{12} = ½\, \varepsilon_{gap} = 2\eta\varepsilon_{JT}$ is the crossover resonance half- splitting energy.

When we use a BT 'classical statistics' for the polaron gas, applying the above approach to the transition probabilities based on nonrelativistic quantum mechanics, sounds normal. Such will not be the case of a squeezed polaron band at low temperature where the use of F-D statistics is mandatory. Extending our realm of normality, the above equations for the transition probabilities should also be replaced by their corresponding relativistic substitutes [14]:

## 4. Reaction rate with quantum statistics

Using specified statistical arguments, we rewrite the reaction rate equation (1) to read

$$k_{12}(T) = (k_B T / h) \sum_\alpha P(E_\alpha)_{specified} W(E_\alpha) (\Delta E_\alpha / k_B T) \quad (31)$$

where $P(E_\alpha)_{specified}$ is given by eq. (15) for F-D, (22) for B-E, and (19) for classical statistics. In particular, eqs. (15) and (19) are ultimately aimed at describing phonon-coupled transitions at a system of squeezed polaron bands. Next, we reproduce the relevant rates from eqs.(15), (22), and (19) at δ = F-D, B-E and classic, respectively, averaging over $n_k$ as in eqs. (16), (18), and (23) and normalizing, as follows:

$$< k_{12}(T)_\delta >_{avnk} = (1 / h) \sum_k < P(\varepsilon_k)_\delta >_{avnk\ norm} W(\varepsilon_k) \Delta\varepsilon_k \quad (32)$$

$$< P(\varepsilon_k)_\delta >_{avnk\ norm} = < P(\varepsilon_k)_\delta >_{avnk} / \sum_k P(\varepsilon_k)_\delta \quad (33)$$

$$< P(\varepsilon_k)_\delta >_{avnk} = 1 / ( \exp[(\varepsilon_k - \mu) / k_B T] + 1 ), (\delta = \text{F-D}) \quad (34)$$

$$< P(\varepsilon_k)_\delta >_{avnk} = 1 / ( \exp[(\varepsilon_k - \mu) / k_B T] - 1 ), (\delta = \text{B-E}) \quad (35)$$

$$< P(\varepsilon_k)_\delta >_{avnk\ norm} = \exp(-\varepsilon_k / k_B T) / \sum_k \exp(-\varepsilon_k / k_B T), (\delta = \text{Classic}) \quad (36)$$

with normalization factors to eqs. (34) and (35), respectively,

$$N_{F\text{-}D} = \ln ( \exp[(\mu - ½\eta\omega) / k_B T] + 1 ) \quad (37)$$

$$N_{B\text{-}E} = \ln ( \exp[(\mu - ½\eta\omega) / k_B T] - 1 ) \quad (38)$$

In what follows, we will compare the rates at δ = F-D, B-E, and classic, as calculated by means of eqs. (32) through (36) at reasonable parameters pertinent to actual systems. In order to do that, we need not calculate the rates themselves. Indeed, since according to eq. (32) various rates therein differing by the way they define the thermal occupation probabilities, we may compare the thermal occupation factors themselves. The vibronic model requires applying a statistics onto a coupled oscillator with eigenenergies $\varepsilon_k = (k + ½) \eta\omega$. We do that and show the results graphically in Figure 1.

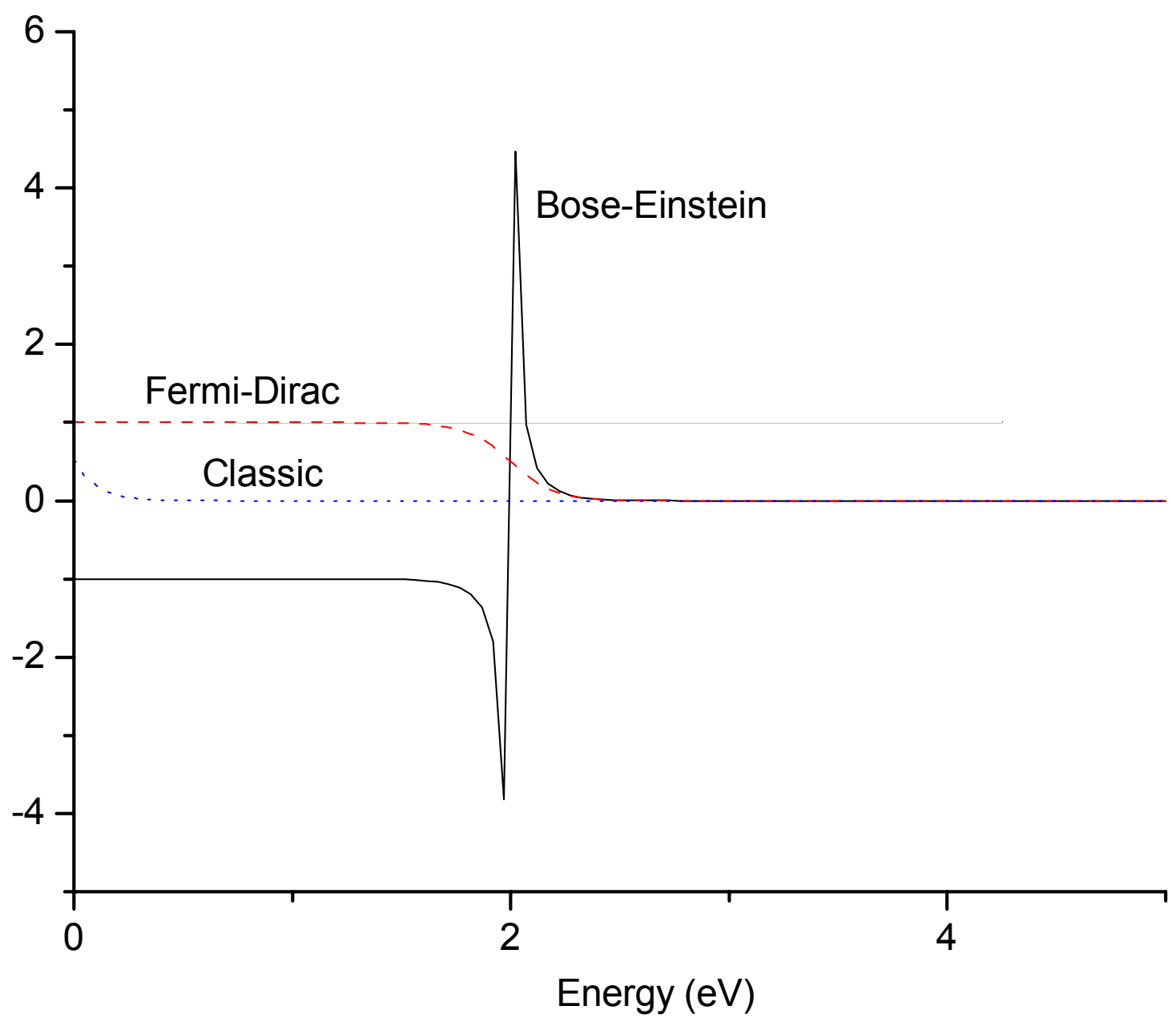

Figure 1

Particle energy distribution by quantum statistics: Average occupation number by Fermi-Dirac statistics from eq. (34) (dashed red) and the average occupation number by Bose-Einstein statistics from eq. (35) (solid black). The Fermi energy is $\mu$ = 2 eV, the temperature $k_B T$ = 0.1 eV, and the phonon quantum $\eta\omega$ = 0.075 eV. (The B-E branch below the Fermi level is unphysical.) For comparison, classic statistics is also shown (dotted blue), as calculated from eq. (36) using the harmonic oscillator model.

We see that imposing the F-D statistics will extend the low-temperature constant rate branch up to energies some twelve percent below the Fermi level just where Boltzmann tail (BT) statistics starts showing up. The polaron gas called "degenerate" within BT, the temperature dependence of its rate is indistinguishable from a classic rate, even the dependency on the Fermi energy cancels out. On the other hand, a B-E rate will duplicate most of the features displayed by the classic and BT rates, though on a different scale.

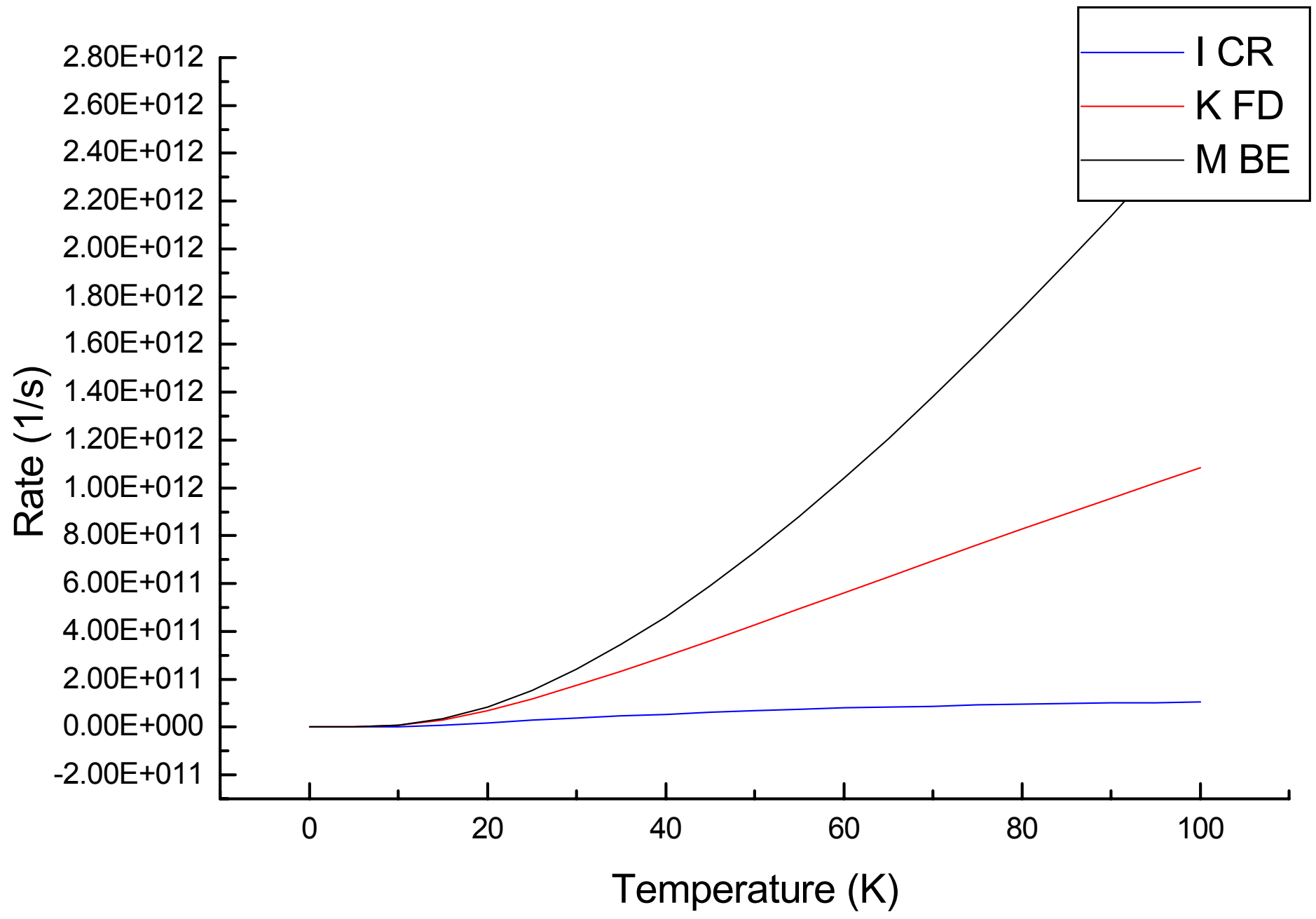

Figure 2

Comparative calculations of quantum statistics reaction rates performed using parameters pertinent to ‘in-plane’ bond polarons in the $La_{2-x}Sr_xCuO_4$ high-$T_c$ superconductor at x = 0.08 (Refeerence 11). From top to bottom: Temperature dependence of a rate with Bose-Einstein (black), Fermi-Dirac (red), and Classic Rate (blue) statistics. The Fermi level was chosen to be at $E_F$ = 1 meV.

Figure 2 shows the result of a rate calculation for bond polarons in a high-$T_c$ superconductor, believed to be a quantum system at low temperatures. For calculating the rates we used the ‘average probabilities’ from equations (16) and (23) for FD and BE, respectively, as well as the normalized probability from eq. (19) for CR. It can be seen that the B-E statistics makes a rate superior to the rates by both F-D and Classic Rate statistics. The quantum statistics thus provide higher rates as far as the absolute magnitudes are concerned. As to the slopes, they are apparently similar in the three cases, which is not surprising in so far as the same microscopic system is involved. These general features should be reckoned with while considering the feasibility of any particular choice of statistics.

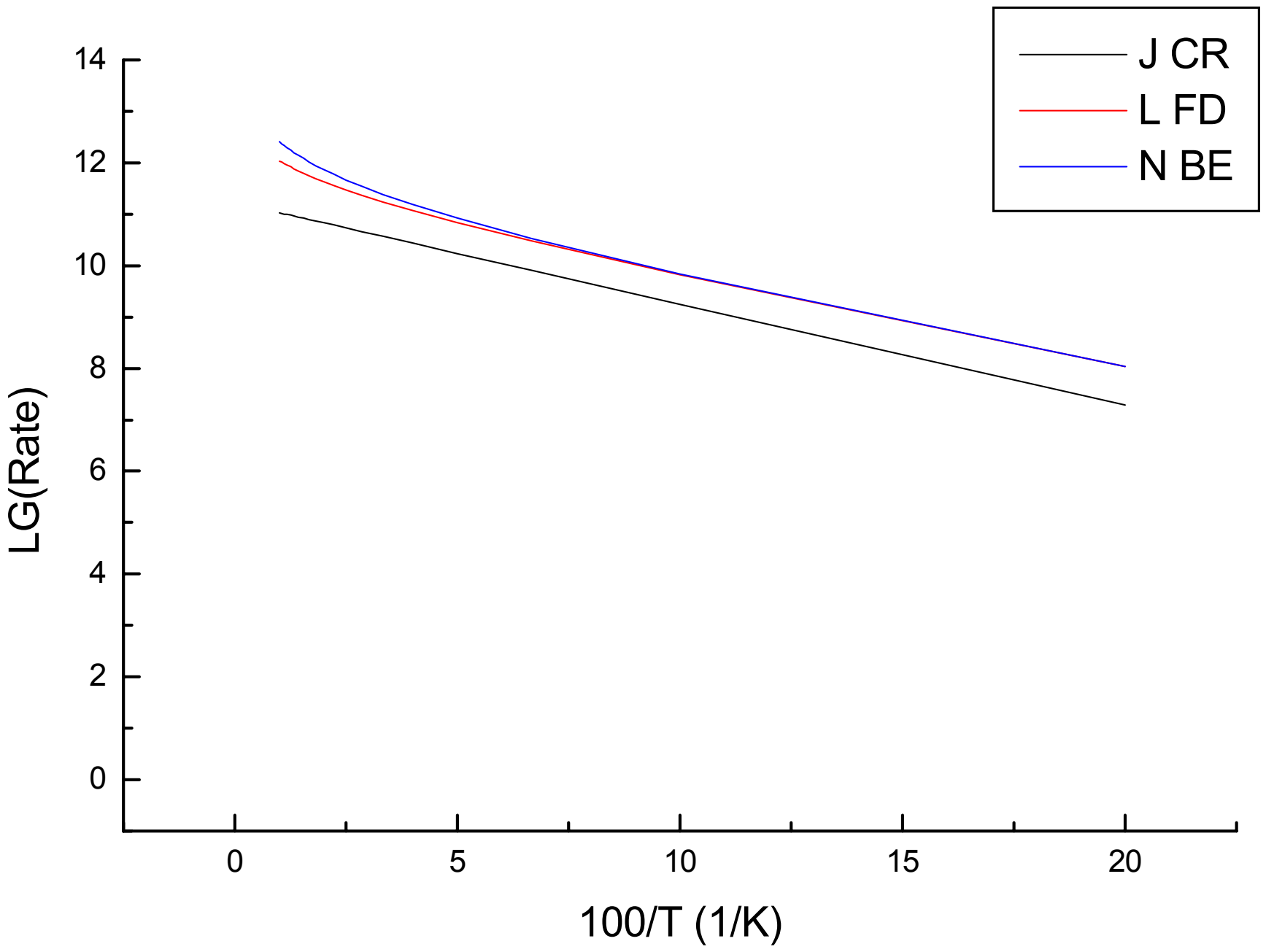

Figure 3

Same as Figure 2 in the Arrhenius temperature plot. From top to bottom: Bose-Einstein (blue), Fermi-Dirac (red), Classic Rate (black).

Using the normalized probabilities from eq. (37) and (38) for FD and BE may be expected to lower the quantum rates somewhat so as to make them more conciliatory to the classical rate.

In conclusion, we have shown that quantum statistics can be introduced in a rate along the same lines as classic statistics and leads to comparable, within an order of magnitude, numerical results.

References

[1] S.G. Christov, *Collision Theory and Statistical Theory of Chemical Reactions*. Lecture Notes in Chemistry #18. (Springer, Berlin, 1980).
[2] P. Hanggi, P. Talkner, M. Borkovec, *Reaction-rate theory: fifty years after Kramers*, Revs. Mod. Phys. **62**, 252-341 (1990).
[3] T. Holstein, Ann. Phys. (N.Y.) **8**, 325 (1959).
[4] T. Holstein, Ann. Phys. (N.Y.) **8**, 343 (1959).
[5] M. Georgiev, Revista Mexicana de Fisica **31**, 221-257 (1985).
[6] M. Georgiev,.J. Inform. Record. Mater. **13**, 75-93, 177-195, 245-256 (1985).

[7] S.G. Christov, Phys. Rev. B **26**, 6918-6935 (1982):
[8] M. Georgiev, A. Gochev, S.G. Christov, A. Kyuldjiev, Phys. Rev. B **26**, 6936-6946 (1982).
[9] P.C. Petrova, M.D. Ivanovich, M. Georgiev, M.S. Mladenova, G. Baldacchini, R.-M. Monterali, U. M. Grassano, and A. Scacco, Proc. 13th Internat. Conference on Defects in Insulating Materials: ICDIM '96, G.E. Matthews and R.T. Williams, eds., Winston-Salem NC, (Materials Science Forum, Volumes **239-241**, 1997) 377-380.
[10] L. Brillouin and M. Parodi, *Propagation des Ondes dans les Milieux Periodiques*, (Dunod, Paris, 1956).
[11] See A.G. Andreev, S.G. Tsintsarska, M.D. Ivanovich, I. Polyanski, M. Georgiev, A.D. Gochev, Central Eur. J. Phys. **2** (2) 89-116 (2004) and references therein.
[12] J.M. Blatt, *Theory of Superconductivity* (Academic, New York, 1964).
[13] A.S. Alexandrov, *Phys. Letters A* **236**, 132 (1997).
[14] M. Tashkova-Doneva, PhD Thesis, Bulgarian Academy of Sciences, Sofia 2000 and references cited therein.